\documentclass[aps,twocolumn,superscriptaddress,showpacs,amssymb,prb,floatfix]{revtex4-1}
\usepackage{graphicx,color}
\usepackage{dcolumn}
\usepackage{amsmath}
\usepackage{amssymb}

\usepackage[dvipdfm]{hyperref}
\hypersetup{colorlinks=true,linkcolor=blue,
  implicit=true,breaklinks=true,pagebackref=true,backref=true,
  bookmarks=true,bookmarksnumbered=true,hyperfootnotes=true,debug=true,
  naturalnames=false,citecolor=blue,pdfview=FitH,pdfstartview=FitH,hyperindex=true}

\newcommand{\slrr}      {$T_1^{-1}$}
\newcommand{\lanio}       {La$_3$Ni$_2$O$_6$}

\begin{document}

\thispagestyle{myheadings}

\title{NMR evidence for spin fluctuations in the bilayer nickelate La$_3$Ni$_2$O$_6$}

\author{N. apRoberts-Warren}
\author{J. Crocker}
\author{A. P. Dioguardi}
\author{K. R. Shirer}
\affiliation{Department of Physics, University of California, Davis, CA 95616, USA}
\author{V. V. Poltavets}
\affiliation{Department of Chemistry, Michigan State University, East Lansing, MI 48824, USA}
\author{M. Greenblatt}
\affiliation{Department of Chemistry and Chemical Biology, Rutgers, The State University of New Jersey, 610 Taylor Road, Piscataway, NJ 08854, USA}
\author{P. Klavins}
\author{N. J. Curro}
\email{curro@physics.ucdavis.edu}
\affiliation{Department of Physics, University of California, Davis, CA 95616, USA}

\date{\today}

\begin{abstract}
We report nuclear magnetic resonance data in the bilayer nickelate La$_3$Ni$_2$O$_6$. This material belongs to a family of low valence nickel oxides with square planar coordination of the Ni ions and is isoelectronic to the high temperature superconducting cuprates.  Although the three layer nickelate compound exhibits a spin-state transition accompanied by antiferromagnetic order, the bilayer material shows no phase transition.  The NMR data in the bilayer {La$_3$Ni$_2$O$_6$} reveal the presence of antiferromagnetic spin fluctuations down to 5K that are surprisingly similar to those present in the {magnetically} ordered trilayer {La$_4$Ni$_3$O$_8$}, suggesting a similar electronic structure in the paramagnetic phase of both compounds.
\end{abstract}


\maketitle

\section{INTRODUCTION}

Low-valence nickel oxides have received attention recently, because of their unusual electronic properties and possible relation to the high temperature superconducting cuprates.\cite{Poltavets2010,PickettLa4Ni3O8PRL}  The Ruddlesden-Popper-based T'-type series La$_{n+1}$Ni$_n$O$_{2n+2}$ is particularly interesting, because it contains sets of $n$ NiO$_2$ infinite planes with an average Ni valence of $(n+1)/n$ and electronic configuration of $d^{9-1/n}$.  This configuration is formally analogous to the CuO$_2$ planes in the cuprates  with a hole doping $\delta=n^{-1}$, therefore these compounds might be expected to exhibit properties similar to the high temperature superconducting cuprates.\cite{anisimov}  Indeed the $n=2$ compound is {isostructural} to superconducting La$_2$CaCu$_2$O$_6$, and the $n=3$ compound exhibits an unusual phase transition to a magnetically ordered state below 105 K.\cite{Poltavets2010}  Recent nuclear magnetic resonance (NMR) experiments in  La$_4$Ni$_3$O$_8$ revealed the presence of antiferromagnetic spin fluctuations that emerge below 160 K, and broadening of the $^{139}$La spectrum associated with static moments in the ordered phase.\cite{ApRoberts-Warren2011}

{Unlike the cuprates, however,} the nature of the phase transition in {La$_4$Ni$_3$O$_8$} appears to be
determined by subtle competitions between multiple interactions, rather than an antiferromagnetic exchange between Ni moments, as in the cuprates. In {the NiO$_2$ planes} the square planar crystal field potential lifts the degeneracy of the Ni $e_g$ orbitals, raising the energy of the  Ni $d_{x^2-y^2}$ {orbital} relative to the $d_{z^2}$ orbital.  The Hund's couplings between the Ni $d$-electrons are of comparable magnitude to the crystal field splitting, so the {ground state configuration is not clear.} {Complicating this local picture is the fact that}  the $n$ $d_{z^2}$ orbitals in the unit cell hybridize along the $c$-direction giving rise to a molecular orbital basis.\cite{PickettLa4Ni3O8PRL} Depending on the relative magnitudes of these  interactions the ground state {in both the $n=2$ (with Ni$^{+1.5}$) and $n=3$ (with Ni$^{+1.33}$) compounds} can be either high-spin ($S=2/3$) and insulating or low-spin and metallic ($S=1/3$).\cite{nickelatespressurePardo,GreenblattElectronicCalculationsNickelatesPRB}  Recent high resolution x-ray diffraction studies in the $n=3$ compound revealed an increase in the in-plane lattice constant and a decrease in the out-of-plane lattice constant below the phase transition. \cite{Ni438PressurePRL} This result suggests that the phase transition is a spin-state transition to a low temperature  high-spin insulating phase.  This interpretation is further supported by high pressure studies that indicate a suppression of the ordering temperature with pressure.\cite{Ni438PressurePRL} In both spin states antiferromagnetic coupling between Ni spins in the plane can occur via a superexchange between the $d_{x^2-y^2}$ orbitals through the oxygen 2p orbitals.  The magnitude of this coupling is suppressed relative {to the Cu-Cu superexchange coupling} in the cuprates, because the energy of the 2p states in {La$_4$Ni$_3$O$_8$ is} much lower than the Fermi energy in the case of the nickelates.\cite{nickelatespressurePardo}

{Earlier NMR results on La$_4$Ni$_3$O$_8$ suggested that the high temperature phase is metallic.\cite{ApRoberts-Warren2011} This result is surprising, because resistivity measurements have suggested insulating behavior.  It is possible, however, that the intrinsic resistivity is obscured by the fact that the measurements were conducted on powder samples.  Theoretical calculations of the band structure have suggested both metallic\cite{Poltavets2010} and insulating\cite{PickettLa4Ni3O8PRL} behavior, and at present the question remains open.}

\begin{figure*}
\includegraphics[width=0.2\linewidth]{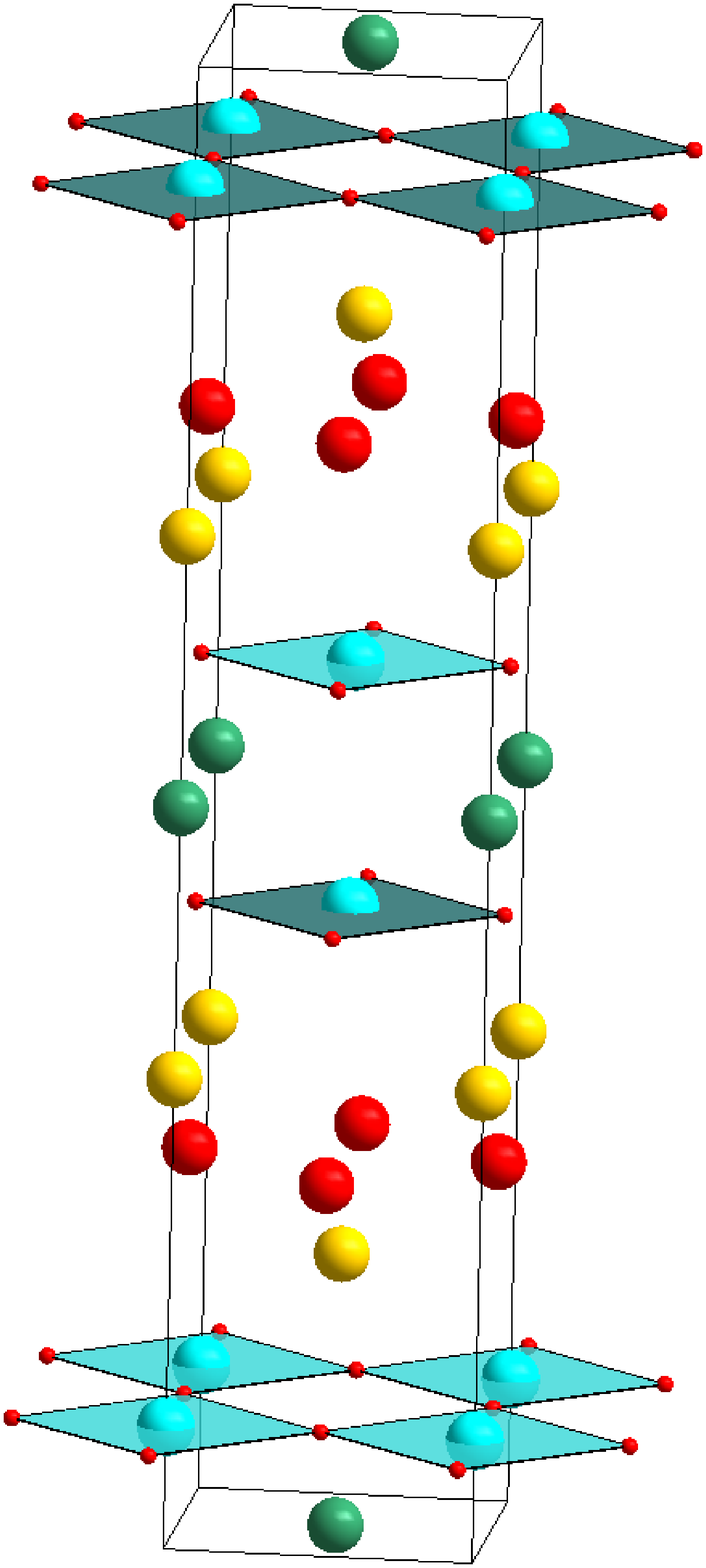}
\includegraphics[width=0.75\linewidth]{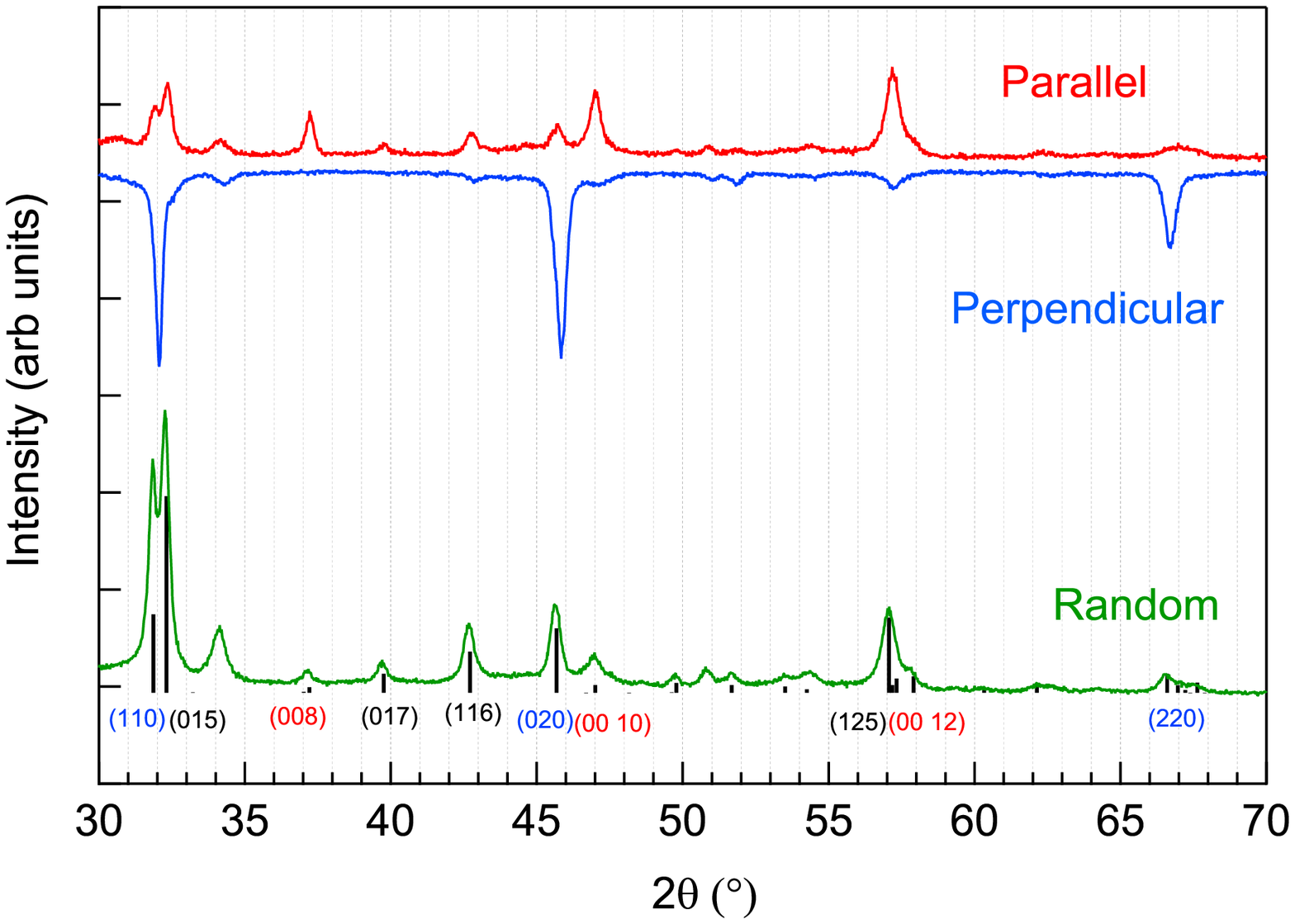}
\caption{(Color Online) {(Left) The structure of \lanio, with Ni (cyan), oxygen (red), La(1) (green) and La(2) (yellow). (Right)} X-ray diffraction pattern for the \lanio\ for the random powder and the aligned sample for the orientation direction aligned parallel and perpendicular to the scattering plane. The data for the perpendicular direction have been inverted for ease of comparison. }
\label{fig:xrays}
\end{figure*}

In contrast to the three-layer compound, \lanio\ shows no phase transition and the spin state and the absence of a phase transition remain open questions. The crystal field potential, the Hund's couplings, and the antiferromagnetic exchange interactions should be comparable to the three-layer material. The primary difference between the two materials is the presence of two or three hybridized $d_{z^2}$ orbitals forming the molecular basis. {There is no obvious reason, therefore, why La$_4$Ni$_3$O$_8$ exhibits long range order, whereas \lanio\ does not. In order to investigate this discrepancy further} we have conducted NMR studies of the $^{139}$La in  \lanio\ to compare the spin fluctuations of the Ni between both materials.  Our results indicate that the spin fluctuations in the two-layer compound are similar to those in the high-temperature phase of the three-layer compound, {which suggest that the increased dimensionality of the three layer materials plays a key role.}

\section{Experiment}

Polycrystalline samples of \lanio\ (tetragonal with space group $I_4/mmm$) were prepared by low temperature reduction of the Ruddlesden-Popper  La$_3$Ni$_2$O$_{7}$ $n=2$ phase as described in Refs. \onlinecite{La326synthesis} and \onlinecite{Poltavets2007}.  The unit cell,  shown in Fig. \ref{fig:xrays}, consists of  infinite bilayers of NiO$_2$ separated by {fluorite La$_2$O$_2$ layers}.  In order to enhance the NMR signal we attempted to form aligned powders by mixing the polycrystalline material with epoxy and curing in a magnetic field, as described in Ref. \onlinecite{ApRoberts-Warren2011}. {In materials with strong magnetic anisotropy the crystallites will tend to align along the axis of greatest susceptibility and are thus fixed in orientation in the epoxy.} To quantify the degree of alignment we measured the powder x-ray diffraction pattern for the alignment direction oriented both within and perpendicular to the scattering plane. As seen in Fig. \ref{fig:xrays} scattering from $(00l)$ planes are nearly absent for the perpendicular orientation, and scattering from $(hk0)$ planes are nearly absent for the in-plane orientation (the structure is shown in Fig. \ref{fig:xrays}).  This result indicates that the crystallites in the powder/epoxy mixture are at least partially aligned. The ratio of the $(00~10)$ reflection intensity to that of the $(020)$ {intensity} is enhanced by a factor of $\sim 9.1\pm 1.4$ in the aligned sample compared to the random powder.\footnote{{Intensities were determined by Gaussian fits, and the background was subtracted.  Error estimates are based upon the best fits to these intensities.}}  By correcting the powder diffraction pattern for preferred orientation using the March-Dollase function, we estimate a March coefficient $G\approx0.38\pm0.05$, which indicates that roughly 70 percent of the crystallites are oriented to within 40$^{\circ}$ of the the c-direction.\cite{leventouri,shaked}  This alignment is slightly less than that in previous studies in La$_4$Ni$_3$O$_{8}$ powders,\cite{ApRoberts-Warren2011} and suggests that the magnetic susceptibility of the bilayer material is less anisotropic at room temperature than the trilayer.

\begin{figure}
\includegraphics[width=\linewidth]{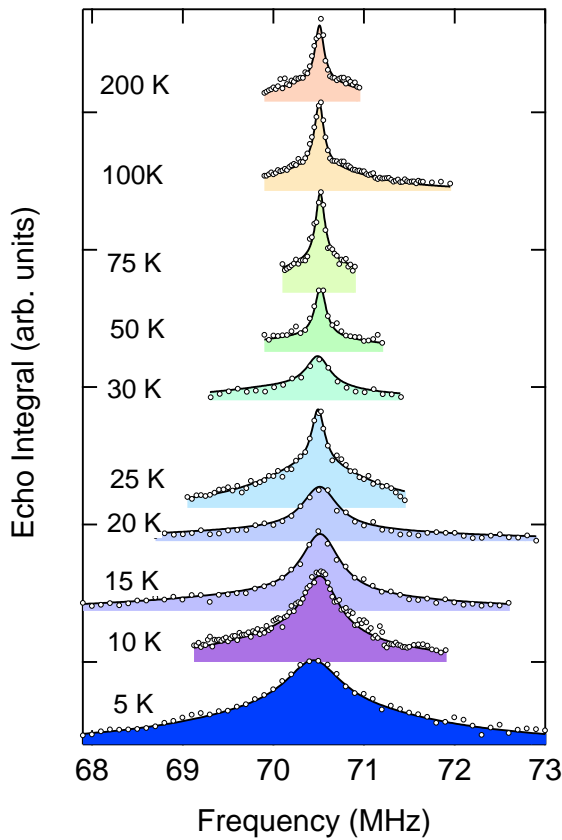}
\caption{(Color Online) NMR spectra of the $^{139}$La in a powder-aligned of La$_3$Ni$_2$O$_6$ measured at several different temperatures at 11.7 T {using a repetition rate on the order of $5T_1$}. The solid lines are fits to two Lorentzians as described in the text.}
\label{fig:spectra}
\end{figure}

In order to investigate the magnetic behavior of this material, we measured the $^{139}$La ($I=7/2$) NMR spectra and relaxation rates as a function of temperature in a field of $H_0=11.7$ T.  The spectra shown in Fig. \ref{fig:spectra} were obtained by adding NMR spin echoes  acquired by the Carr-Purcell-Meiboom-Gill sequence (CPMG) and integrating the  intensity as a function of frequency.   There are two different La sites in the structure: La(1) located between the NiO$_2$ bilayers and La(2) located in the {fluorite La$_2$O$_2$ layers} (see Fig. \ref{fig:xrays}).  Both sites have axial symmetry and the La nuclei are sensitive to the local electric field gradient (EFG). The resonances are determined from the nuclear  {spin} Hamiltonian:  $\mathcal{H} = \gamma\hbar\mathbf{H}_0\cdot(1 + \mathbf{K})\cdot\mathbf{\hat{I}} + (h\nu_Q/6)(3\hat{I}_c^2 - \hat{I}^2)$, where $\gamma$ is the gyromagnetic ratio, $K$ is the magnetic shift, $\nu_Q= 3eQV_{cc}/2I(2I-1)h$ is the quadrupolar frequency, $e$ is the electron charge, $Q$ is the quadrupolar moment, and $V_{cc}$ is the component of the EFG tensor corresponding to the tetragonal $c$ axis of the unit cell. \cite{CPSbook} For an applied field along the $c$ direction the resonance frequency for each site is given by: $\omega_n = \gamma H_0(1+K) + n\nu_Q$, where $n=-3,\cdots+3$.  The NMR spectra reveal no distinct quadrupolar satellites, but do exhibit both broad and narrow features.    It is likely that although the sample is partially aligned, the satellite features remain broadened due to a distribution of orientations.  {This broadening is not due to disorder in the oxygen stoichiometry.  There is no room in this crystal structure to accommodate extra oxygens, and oxygen vacancies would lead to unphysical values of the Ni coordination. We estimate that the oxygen concentration is correct to within less than one percent.}

\begin{figure}
\includegraphics[width=\linewidth]{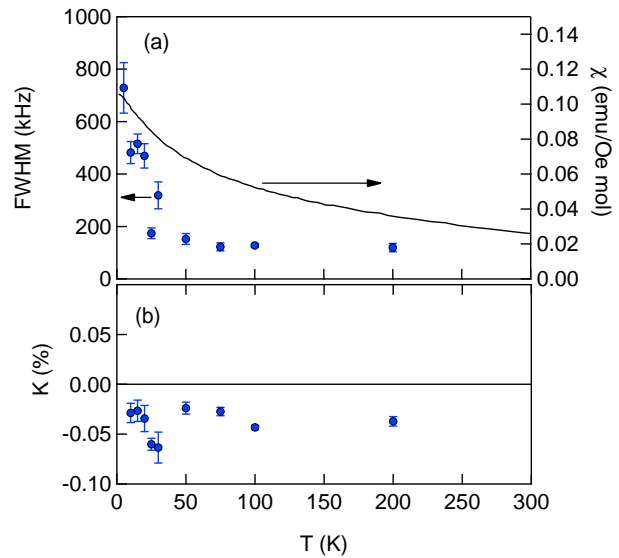}
\caption{(Color Online) The full-width-half-maximum (a) and magnetic shift (b) of the La resonance as a function of temperature.  The solid line in (a) is the magnetic susceptibility measured {on a random powder} at 1 T.}
\label{fig:width}
\end{figure}

We fit the spectra to the sum of two Lorentzians with different widths and centers of gravity. The  full width at half maximum (FWHM) and the magnetic shift, $K$, of the narrow feature are shown in Figs. \ref{fig:width}a and \ref{fig:width}b.  The narrow feature probably represents the La(2) sites (in the {La$_2$O$_2$ layers}). {We estimate that both $\nu_Q(1)$ and $\nu_Q(2)$ are less than 10 kHz in order to explain the spectrum.} Estimates of the EFGs using Wien2K code yield  $\nu_Q(1) = 5$ kHz and $\nu_Q(2) = 3$ kHz for the La(1) and La(2) sites, respectively. These values are too small to resolve, therefore the spectra must consist of signal from both sites. {In the La$_4$Ni$_3$O$_{8}$   $\nu_Q(1) = 1.35$ MHz and $\nu_Q(2) \lesssim$ 20kHz.\cite{ApRoberts-Warren2011}, and in LaNiO$_3$ the La EFG has $\nu_Q = 1.1$ MHz.\cite{SakaiLaNi03NMR}  It is unclear why the EFG at the La(1) site in the \lanio\ is reduced relative to that in the three-layer compound. Magic angle spinning may be the best technique to extract these quantities with greater precision.}

{As seen in Figs. \ref{fig:spectra} and \ref{fig:width}} there is a clear broadening of the spectra with decreasing temperature.  The origin  of this effect is unclear.  It is possible that since the magnetic susceptibility (solid line in Fig. \ref{fig:width}a) is strongly temperature dependent a distribution of demagnetization fields of each crystallite can result in an overall temperature dependent broadening as seen in Fig. \ref{fig:spectra}.  However the temperature dependence of the FWHM does not appear to scale with $\chi$.  {Furthermore, a field-swept spectra acquired at constant frequency of 44.95 MHz at 10 K (not shown) reveals a FWHM that is a factor of 2.3 larger than the FWHM measured at 11.7 T.  The reason for this decrease in linewidth with increasing field is unknown, but suggests that the origin of the broadening is magnetic, rather than quadrupolar in nature.}

The magnetic shift, $K$, appears to be negative and temperature independent {above 50K}, {which suggests that the La site} experiences only a weak hyperfine interaction to the Ni spins and is probably dominated by a temperature independent chemical shift term. The unusual {step in $K$} around 30 K appears to correlate with the sharp upturn in the FWHM and the onset of enhanced spin fluctuations measured by spin lattice relaxation (see below). There is no corresponding feature evident in the susceptibility {(Fig. \ref{fig:width}(a))} or the specific heat.\cite{Poltavets2010}

\begin{figure}
\includegraphics[width=\linewidth]{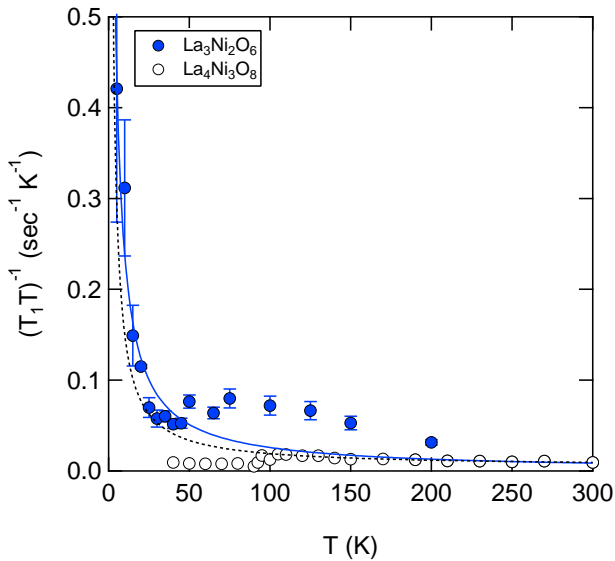}
\caption{(Color Online) (a) $(T_1T)^{-1}$ versus temperature measured at the central peak of the La spectrum for the two-layer \lanio\ ($\bullet$) and the three layer La$_3$Ni$_4$O$_8$ ($\circ$) materials. The solid (dotted) line is a fit to the two(three)-layer data.  Data for the La$_3$Ni$_4$O$_8$  is reproduced from Ref. \onlinecite{ApRoberts-Warren2011}.}
\label{fig:t1}
\end{figure}

\begin{figure}
\includegraphics[width=\linewidth]{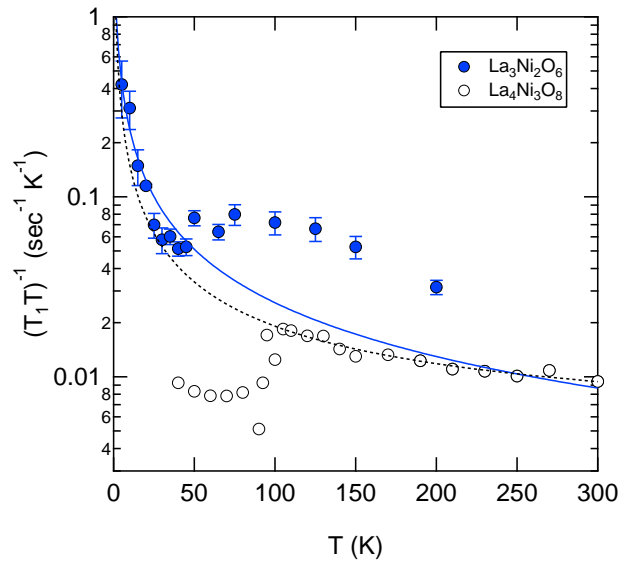}
\caption{(Color Online) (a) $(T_1T)^{-1}$ versus temperature  as in Fig. \ref{fig:t1} plotted on a semilog scale. {The solid and dotted lines are fits as described in the text.}}
\label{fig:t1log}
\end{figure}

The spin lattice relaxation rate, \slrr, was measured by spin echo inversion recovery at the peak of the spectra.  For a single component of relaxation {driven by magnetic fluctuations} measured at the central transition, the magnetization should recover as $M(t) = M_0[1-f\phi(t/T_1)]$, where:\cite{Narathrecovery}
\begin{equation}
\label{eqn:invrec}
\phi(t) = \frac{1225}{1716} e^{-{28 t}}+\frac{75}{264} e^{-{15
   t}}+\frac{3}{44} e^{-{6t}}+\frac{1}{84}{e^{-{t}}}.
\end{equation}
The temperature dependence of $(T_1T)^{-1}$ is shown in Figs. \ref{fig:t1} and \ref{fig:t1log}.
The fits to the data are not particularly good at low temperatures, which is likely due to the fact that the spectra consist of both La sites and several different quadrupolar transitions may be excited simultaneously. Nevertheless this function gives an estimate of the relaxation rate and enables us to discern the temperature dependent trend.  We also fit the magnetization recovery to a stretched exponential form; this procedure gave a better fit to the magnetization recovery, but with greater scatter in the temperature dependent data. Similar results were found using  two relaxation components as in Ref. \onlinecite{ApRoberts-Warren2011}, however the qualitative trend evident in Fig. \ref{fig:t1} did not change using either stretched exponentials or two component fits.  The data shown in Figs. \ref{fig:t1} and \ref{fig:t1log} were fit using a single unstretched component, and  reveal a slight peak at 80K, followed by a shallow minimum at 40K, and a sharp increase below. The increase coincides with the increase in the linewidth and drop in $K$ evident in Fig. \ref{fig:width}.

\section{Results and Discussion}

The surprising increase with decreasing temperature observed in $(T_1T)^{-1}$ and in the NMR linewidth suggest the presence of critical spin fluctuations.  Similar behavior was reported in the La$_4$Ni$_3$O$_8$ material, in which 2D spin fluctuations appeared to diverge close to $T=0$, but were cut off by the emergence of long range order at 105 K.\cite{ApRoberts-Warren2011}  This result is surprising, given that La$_3$Ni$_2$O$_6$ has no long range {magnetic} order.  In order to quantify the spin fluctuations we fit the data to the expression $(T_1T)^{-1} = a + b/(T-T_0)$ {as used for the three-layer compound,\cite{ApRoberts-Warren2011}} where the first term represents a contribution from Korringa scattering from spin-flip scattering from quasiparticles, and the second term represents a contribution from antiferromagnetic spin fluctuations.  The best fit to the data below 50 K yields the values {$a=0.0\pm0.1$ s$^{-1}$K$^{-1}$, $b=2.6\pm1.2$ s$^{-1}$ and $T_0 = -1\pm3$ K}. These parameters are close to those reported previously for the La$_4$Ni$_3$O$_8$ ($a=0.0045$ s$^{-1}$K$^{-1}$, $b=1.46$ s$^{-1}$ and $T_0 = 0$ K), reproduced in Fig. \ref{fig:t1}.  These results suggest that both compounds exhibit similar 2D fluctuations, but in the trilayer compound the material undergoes a phase transition at 105 K. Since this transition is not present in the bilayer compound the spin fluctuations can continue to develop down to low temperature. The data suggest that if the three-layer compound did not undergo the phase transition then the spin fluctuations would be remarkably similar to those of the two-layer compound. {The vanishing Korringa term suggests that \lanio\ is insulating, and contrasts with the observations in La$_4$Ni$_3$O$_8$.  However fitting $(T_1T)^{-1}$ over the entire temperature range gives a finite value ($a=0.024\pm 0.015$ s$^{-1}$K$^{-1}$). Detailed measurements of \slrr\ for much higher temperatures (well above room temperature)  may be necessary to resolve whether the the temperature independent contribution (the $a$ term) persists when the spin fluctuation contribution (the $b$ term) is suppressed.}

The picture that emerges is that both compounds have the same electronic configuration in the high temperature paramagnetic phase.  The presence of  spin fluctuations suggest antiferromagnetic coupling between the Ni moments, with an exchange coupling on the order of 100 K.  If the spin and/or electronic configuration were different between the two materials, then it is unlikely that the temperature dependence of \slrr\ would be so similar.  Since these spin fluctuations are primarily two-dimensional in the NiO$_2$ plane, no long range order develops without some other interaction present.\cite{MerminWagner}  In the case of the three layer La$_4$Ni$_3$O$_8$ the spin-state transition at 105 K changes the ground state electronic configuration, which also enables the material to undergo long-range magnetic order. In the two-layer case no such change in the electronic configuration takes place, and therefore the antiferromagnetic spin fluctuations continue to grow down to the lowest temperatures measured.  {These results suggest that the enhanced c$-$axis coupling in the three layer compound is responsible for the long range order and the stabilization of the low temperature state. Surprisingly, however, the suppression of the long-range order by reduced dimensionality  in these compounds appears to contrast with that observed in LaNiO$_3$ heterostructures, where magnetic order is stabilized in alternating layers of LaNiO$_3$/LaAlO$_3$ superlattices.\cite{LaNiO3heterostructuresScience2011} In the latter, however, the ordering is believed to arise because of the enhanced nesting of the Fermi surface.  If \lanio\ is indeed insulating, as our \slrr\ results suggest, then such an effect would not be present.}

{In summary, we have measure the $^{139}$La NMR spectra and \slrr\ in the two-layer \lanio\ Ruddlesden-Popper compound and found evidence for spin fluctuations similar to those observed in the three-layer analog. Coupling along the $c$-axis of the latter leads to long range order in La$_4$Ni$_3$O$_8$, but it appears to be absent in \lanio. }

\section{Acknowledgements}

We thank G. Kotliar,  C. Panagopoulos,  {W. Pickett} and R. Stern for stimulating discussions. Work at UC Davis was supported by the the NSF under Grant No.\ DMR-1005393, VP was supported by NSF-DMR-1206718, and MG was supported by NSF-DMR-0966829.


\bibliography{CurroBibliography}

\end{document}